\documentclass[journal=jpclcd,manuscript=letter]{achemso}

\usepackage[version=3]{mhchem} % Formula subscripts using \ce{}
\usepackage[T1]{fontenc}       % Use modern font encodings
\usepackage{gensymb}
\usepackage{graphicx}

\author{Robert Malinowski}
\affiliation[UCL]
{Department of Chemistry, University College London, 20 Gordon Street, London WC1H 0AJ, United Kingdom}
\author{Giovanni Volpe}
\affiliation[University of Gothenburg]
{Department of Physics, University of Gothenburg, 41296 Gothenburg, Sweden}
\author{Ivan P. Parkin}
\affiliation[UCL]
{Department of Chemistry, University College London, 20 Gordon Street, London WC1H 0AJ, United Kingdom}
\author{Giorgio Volpe}
\email{g.volpe@ucl.ac.uk}
\affiliation[UCL]
{Department of Chemistry, University College London, 20 Gordon Street, London WC1H 0AJ, United Kingdom}

\title[Malinowski2017]
  {Dynamic Control of Particle Deposition in Evaporating Droplets by an External Point Source of Vapor}

\begin{document}


\begin{abstract}
The deposition of particles on a surface by an evaporating sessile droplet is important for phenomena as diverse as printing, thin-film deposition and self-assembly. The shape of the final deposit depends on the flows
within the droplet during evaporation. These flows are typically determined at the onset of the process by the intrinsic physical, chemical and geometrical properties of the droplet and its environment. Here, we demonstrate deterministic emergence and real-time control of Marangoni flows within the evaporating droplet by an external point-source of vapor. By varying the source location, we can modulate these flows in space and time to pattern colloids on surfaces in a controllable manner.
\end{abstract}


% \section{Introduction}

When a liquid droplet containing small solid particles dries on a surface, it leaves behind a characteristic stain of deposited material that is often in the shape of a ring. The mechanism that leads to this non-uniform deposition, known as the ``coffee ring effect'', arises in a wide range of situations where the contact line of the evaporating droplet is pinned \cite{Deegan1997}: during the drying process, faster evaporation at the droplet's edge induces a radial capillary flow that replenishes the liquid evaporating there with liquid from the droplet's center; the same flow carries suspended or dissolved material to the edge, where it forms a ring-shaped deposit \cite{Deegan1997,Deegan2000,Deegan2000_2}. Recently, these non-equilibrium dynamics, and their control, have garnered a lot of attention because of their fundamental interest and potential applications \cite{Yodh2013,Larson2013}. The patterns left by a drying droplet on a surface are, for example, of interest for several technological applications, such as printing, coating, thin-film deposition and self-assembly \cite{Wei2012}.

Although the coffee ring effect is ubiquitous, recent work has shown that its dynamics can be altered and even reversed, for example, by varying the size and shape of the suspended particles \cite{Weon2010,Yunker2011,Araujo2015}, by introducing surfactants \cite{Still2012, Sempels2013}, by inducing temperature gradients \cite{Hu2006,Soltman2008,Li2015}, by changing solvent composition \cite{Park2006}, by exposing the droplet to a controlled homogenous atmosphere \cite{Majumder2012} or by controlling pinning and contact angles \cite{Nguyen2013,Li2013,Zhang2014}. Several of these factors, in particular, can counteract the outward capillary flows by introducing surface tension gradients in the evaporating droplet that generate recirculating flows, known as Marangoni eddies \cite{Hu2005,Hu2006,Still2012}. Surfactants have proven to be simple but effective additives to generate stable Marangoni flows, although they are often left in the final stain after evaporation \cite{Still2012,Sempels2013}. Alternatively, temperature-induced Marangoni flows can be generated by heating the substrate or the upper surface of the droplet \cite{Soltman2008,Thokchom2014,Li2015}. Finally, droplets made of binary mixtures can improve the shape of the final deposit \cite{Park2006}; however these mixtures can also lead to uneven evaporation processes \cite{Rowan2000,Sefiane2003} and to the emergence of chaotic Marangoni flows \cite{Christy2011,Kim2016}. 

So far, all the approaches proposed to control the dynamics of evaporating droplets rely on altering the intrinsic physical, chemical and geometrical properties of the droplet, of its substrate or of its atmosphere at the very onset of the evaporation process. Once these initial conditions are set, there is little real-time control over the emergence and generation of the flows within the droplet, and thus on the final deposit of the material in it. Only very recently, a degree of local control over Marangoni flows has been demonstrated using laser radiation alone\cite{Ta2016} or in combination with light-activated surfactants \cite{Varanakkottu2016}. 

Here, we propose a novel mechanism to generate and control Marangoni flows within an evaporating sessile droplet in a deterministic and dynamic way. We use an external point source of vapor to induce a local change in surface tension on the droplet's upper surface, thus allowing the real-time reshaping of the flows within it without altering its temperature and with minimal change in its composition. We further corroborate our experimental observations with simple scaling arguments. Finally, we show how both the onset and strength of this mechanism can be accurately modulated in space and time to pattern a surface with controllable deposits of colloids. 

\begin{figure} 
\includegraphics[width=\linewidth]{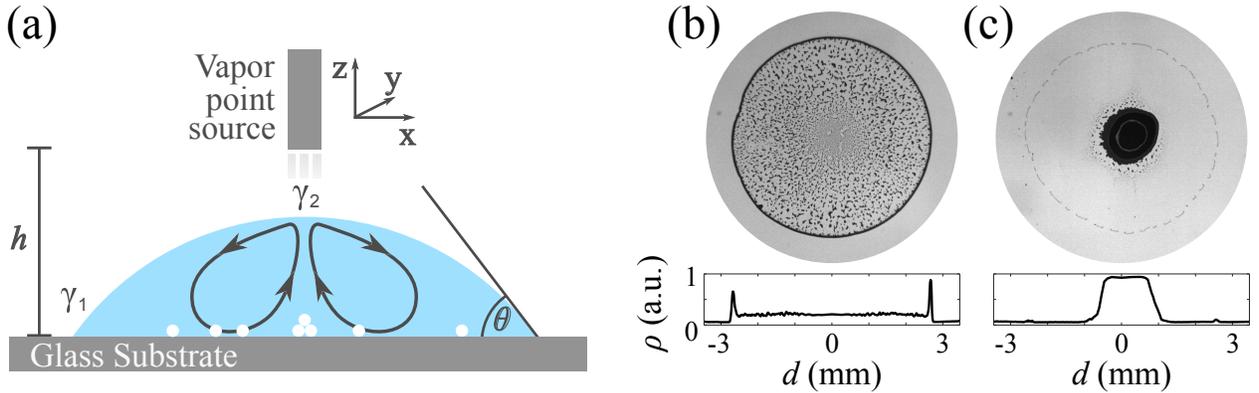}
\caption{\label{fig1} \textbf{Effect of a vapor point source on the final deposit of an evaporating sessile droplet.} (a) Schematic representation of our experimental configuration (not to scale): a sessile droplet containing monodisperse silica particles (white circles) is left to evaporate on a glass substrate (contact angle $\theta$) under a needle containing dry ethanol. The needle is mounted on a three-axis micrometric stage to translate it with respect to the droplet's center and, during the evaporation, it is kept at a controllable distance $h$ from the substrate. When the needle is positioned close to the droplet's upper surface, ethanol vapor induces recirculating Marangoni flows (solid lines) due to a local reduction in surface tension $\gamma$ ($\gamma_1 > \gamma_2$). The arrows on the lines show the direction of the flows. (b-c) Final deposits after evaporation (b) in the absence and (c) in the presence of ethanol in the needle ($h = 2 \, {\rm mm}$). The histograms at the bottom show the corresponding density profile $\rho$ of the deposit along one droplet's diameter as calculated from the image inverted gray scale; to improve the signal-to-noise ratio, these profiles are averaged along the angular coordinate.}
\end{figure}

To observe the effect of a vapor point source on the evaporation dynamics of a sessile droplet, we implemented the experimental configuration depicted schematically in Fig.~\ref{fig1}(a) within an environmental chamber with controlled temperature, $T = 25 \pm 0.2\degree \, {\rm C}$, and relative humidity, $RH = 45\pm5\%$ (see the Supporting Information). Unless otherwise stated, all experiments were performed by depositing a 1-${\rm \mu}$L droplet (radius $R = 2.7 \pm 0.2\, {\rm mm}$) of a 1-wt\% water suspension of 2-${\rm \mu}$m monodisperse silica particles (density $\rho_{\rm Si} = 1850 \, {\rm kg \, m}^{-3}$, sedimentation velocity in water $v_{\rm s} = 2.1 \, {\rm \mu m \, s^{-1}}$) on a clean glass slide (contact angle $\theta \leq 5 \degree$). A needle of inner radius $r_0 = 210 \, {\rm \mu m}$ containing 10 ${\rm \mu}$L of dry ethanol was then positioned above the droplet to provide a constant vapor concentration near the droplet's surface during the whole evaporation experiment (total duration $t_{\rm f} = 250 \pm 20\, {\rm s}$). The needle was mounted on a three-axis micrometric stage to guarantee the possibility of carefully positioning and translating the point source with respect to the droplet in all directions (see the Supporting Information). The evaporation process was then recorded at $10\, {\rm fps}$ (frames per second) with low magnification with a CMOS camera mounted on an inverted microscope with the possibility of switching between bright- and dark-field illumination. Because of the presence of particles in the fluid, the contact line typically remained pinned throughout the duration of the evaporation experiments. 

In the absence of ethanol in the needle, as can be seen in Fig.~\ref{fig1}(b) and in Supporting Movie 1, the coffee ring effect is unperturbed by our system and standard ring-shaped deposits are left after the evaporation process as a consequence of capillary flows \cite{Deegan1997,Deegan2000,Deegan2000_2}. This is no longer the case when ethanol vapor saturates the atmosphere within the needle (vapor pressure $P_{\rm EtOH} = 7.83 \, {\rm Pa}$ at $T = 25 \, \degree {\rm C}$) and diffuses from there towards the droplet's surface, where it induces a local decrease in surface tension $\gamma$, i.e. $\gamma_1 > \gamma_2$  (Fig.~\ref{fig1}(a)) \cite{Vazquez1995}. This difference in surface tension between the top and the edge of the droplet drives the formation of recirculating Marangoni eddies towards the areas of higher $\gamma$ \cite{Still2012}, corresponding to the edge in our case. As can be seen in Fig.~\ref{fig1}(c) and in Supporting Movie 2, when the distance between the vapor point source and the substrate is $h = 2 \, {\rm mm}$, these flows are already strong enough to counteract the coffee-ring effect and to accumulate the suspended particles in a narrow area around the flow stagnation point in the middle of the droplet. For a given height (e.g. $h = 2 \, {\rm mm}$), the shape and size of this central spot depend on the inner radius of the needle $r_0$ (Supporting Fig. S1(a)). If $r_0$ is too big ($r_0 = 640 \, {\rm \mu m}$, approximately a quarter of the droplet's basal radius), the final pattern is strongly distorted due to an excess of ethanol vapor around the droplet. Reducing $r_0$ (from $r_0 = 350 \, {\rm \mu m}$ to $r_0 = 150 \, {\rm \mu m}$) makes the evaporation more controllable, and the corresponding weakening of the Marangoni flows for decreasing $r_0$ progressively makes the final stain smaller and more regular in shape (Supporting Fig. S1(b-c)) while, at the same time, increasing the amount of particles deposited between the central spot and the edge (Supporting Fig. S1(b)). A further reduction in $r_0$ ($r_0 = 80 \, {\rm \mu m}$) makes the influence of the point source of vapor negligible, unless the needle is brought closer to the droplet's upper surface (Supporting Fig. S1(a)).

To a first approximation, the whole process can be understood in terms of a simplified analytical model, where the transient nature of the evaporation is neglected and a steady state is assumed for the diffusion of ethanol vapor from the needle towards the droplet. In this case, the steady-state Poisson diffusion equation can be solved for a spherical source to obtain the concentration of ethanol in air, $c_{\rm A}$, at the upper surface of the droplet as a function of the radial coordinate $r$:
\begin{equation}\label{eq1}
c_{\rm A}(h,r) =\frac{1}{2}\frac{c_0 r_0}{\sqrt{(h-h_D)^2+r^2}},
\end{equation}
where $c_0 = 0.14 \, {\rm Kg \, m^{-3}}$ is the concentration of ethanol at the needle's tip (corresponding to $P_{\rm EtOH}$), $h_{\rm D} = 85 \pm 10\, {\rm \mu m}$ is the droplet's height and, given the small value of $h_{\rm D}$, we have assumed a thin-wedge geometry to simplify the calculations \cite{Nikolov2002}. The prefactor $1/2$ accounts for the fact that the source can only emit in the lower half space in our experimental configuration. Because of its very fast adsorption dynamics \cite{WilsonJPhysChemB1997}, the concentration of ethanol in water, $c_{\rm W}$, is approximately the same as in air, and, provided that the mass percentage of ethanol in water is small ($< 1 \%$) as in our experiments (Supporting Fig. S2), the local surface tension $\gamma(r)$ on the droplet's surface depends linearly on $c_{\rm W}$, so that
% \begin{widetext}
\begin{equation}\label{eq2}
\gamma(h,r) = \gamma_{\rm W} - \beta c_{{\rm W}}(h,r)  = \gamma_{\rm W} - \frac{\beta}{2} \frac{ c_0 r_0}{\sqrt{(h-h_D)^2+r^2}},
\end{equation}
% \end{widetext}
where $\gamma_{\rm W} = 7.2 \cdot 10^{-4} \, {\rm Nm^{-1}}$ is the surface tension of pure water at $T = 25 \degree {\rm C}$, and $\beta = 3.26 \cdot 10^{-4} \, {\rm m^3s^{-2}}$ is a proportionality constant \cite{Vazquez1995}. In order to quantify an upper bound for the concentration of ethanol within the droplets in the presence of the point source of vapor, we monitored how the contact line of 1-${\rm \mu}$L droplets spreads during evaporation on a clean glass slide for water-ethanol binary mixtures at increasing ethanol concentrations. Initial spreading of the droplets, beyond what can be typically observed in the presence of the point source of vapor, only becomes evident above ${\rm [EtOH]} \geq 0.15 \, {\rm v/v}\%$ corresponding to a mass percentage in water of approximately $0.12$ (Supporting Fig. S2). This value represents an upper bound for the change in composition of the droplet induced by the point source of vapor in a typical experiment. Interestingly, below this threshold, in the initial phases of the evaporation, the presence of ethanol in the binary mixture droplet drives some recirculating flows that start to accumulate suspended particles at the center of the droplet (Supporting Fig. S2(b) and Supporting Movie 3). These flows are similar to those we observe in the presence of the point source. Their effect is however short-lived and quickly outperformed by capillary flows as ethanol evaporates. The presence of an external point source of vapor as in Fig.~\ref{fig1} instead establishes a balance between absorbed and evaporated ethanol that allows the recirculating flows to last for the entire evaporation process.

\begin{figure}
\includegraphics[width=\linewidth]{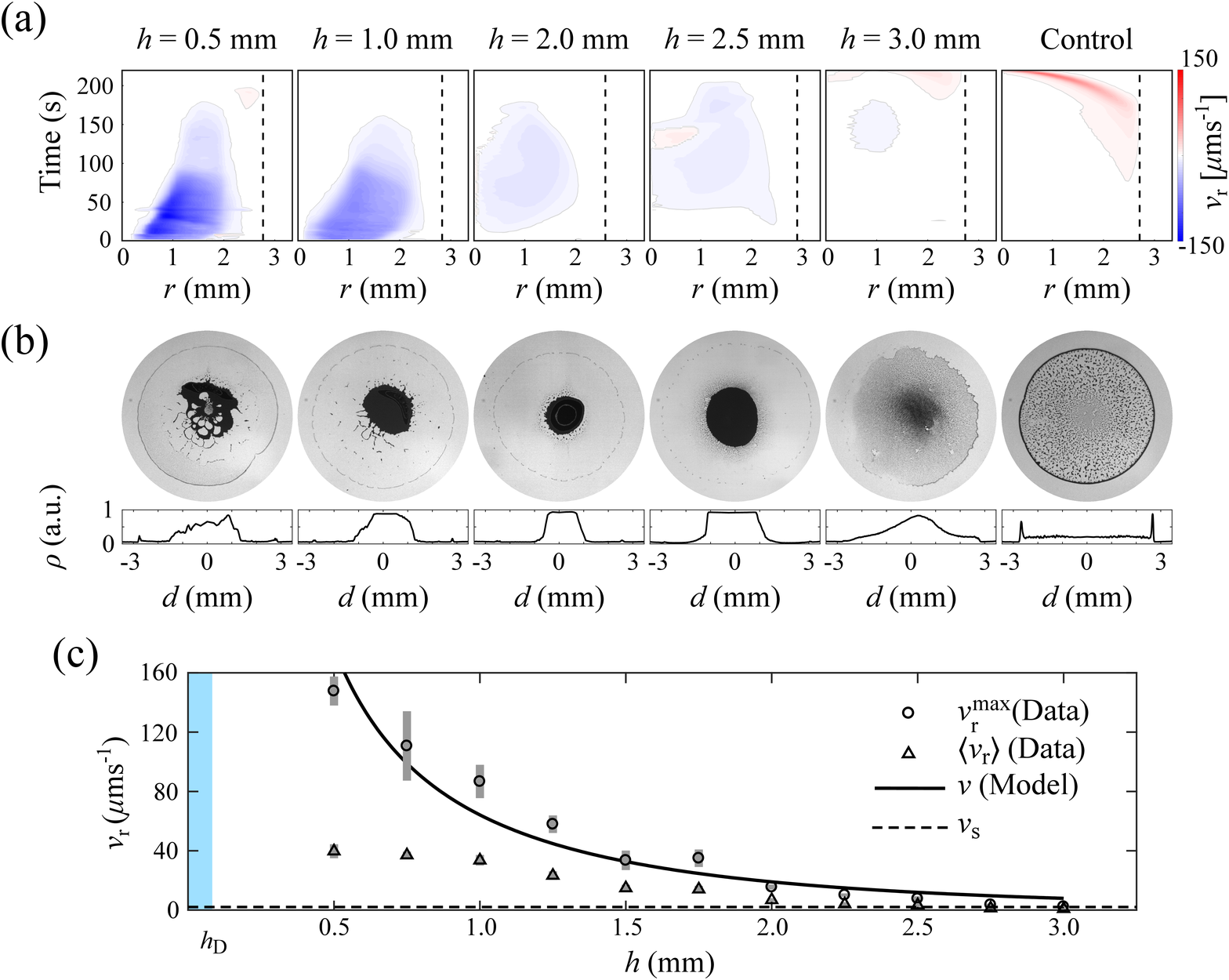}
\caption{\label{fig2} \textbf{Strength of the Marangoni flows and its dependence on the droplet's distance from the vapor point source.} (a) Maps of the radial component, $v_{\rm r}$, of the flow velocity vectors within the droplet near the substrate for different values of $h$ as a function of the radial position $r$  and the time from the start of the evaporation process. Every value is averaged along the angular coordinate. The vertical dashed lines represent the edge of each droplets under observation. (b) Final deposits after evaporation for the droplets of the flow velocity maps in (a). The bottom histograms show the corresponding deposit's density profile $\rho$ (averaged along the angular coordinate) along one droplet's diameter (calculated as in Fig.~\ref{fig1}). (c) Maximum measured radial velocity $v_{\rm r}^{\rm max}$ (circles), average measured radial velocity $\langle v_{\rm r} \rangle$ (triangles) and analytically predicted velocity $v$ (solid line, Eq. \ref{eq3} ) of the Marangoni flows as a function of $h$; the experimental values are averaged over at least 3 different droplets. The horizontal dashed line represents the microparticles' sedimentation velocity $v_{\rm s}$. The gray bars show one standard deviation around the mean values. The blue shaded vertical area indicates the average height of a droplet.}
\end{figure}

Equation \ref{eq2} shows how $\gamma$ is lowered more prominently at the top of the droplet ($r =0$) rather than at its edges ($r = R$), and how this decrease is stronger for increasing values of the needle's inner diameter $r_0$ and decreasing values of the distance $h$. This difference in surface tension drives Marangoni flows that coast the droplet's surface from its center towards it edges and that are then recirculated along the substrate to the top through the droplet's center \cite{Still2012}. This directionality is confirmed by the measurements in Fig.~\ref{fig2}(a), which show the radial component, $v_{\rm r}$, of the flow velocity vectors near the substrate for different values of $h$ as a function of the radial position $r$  and the time from the start of the evaporation. To obtain these velocity vectors, we analyzed videos of evaporating droplets with particle image velocimetry (PIV) \cite{Willert1991}. First, the videos' contrast was enhanced using contrast limited adaptive histogram equalization (CLAHE), then PIV was performed on the enhanced videos using PIVlab with an FFT deformation method applied sequentially with 64$\times$64, 32$\times$32 and 16$\times$16-pixel interrogation windows to improve accuracy \cite{ThielickeJOpRS2014}. Finally, a median filter (5$\times$5 vectors) was applied to the obtained velocity vectors for validation; the vectors were then averaged over 2-s time intervals to further improve the signal-to-noise ratio. 

As predicted by Eq.~\ref{eq2} and demonstrated by the data in Fig.~\ref{fig2}(a), the strength of these flows can be reduced by increasing $h$ until, for long distances ($h \geq 3 \, {\rm mm}$), the standard coffee ring effect takes over very weak Marangoni flows due to its characteristic strengthening towards the end of the evaporation process \cite{Marin2011}. As a control, we also verified that only the outwards capillary flows of the coffee ring effect are present in absence of ethanol. Fig.~\ref{fig2}(b) shows the final deposits corresponding to the flow velocity maps of Fig.~\ref{fig2}(a): here, the strengthening of the Marangoni flows with decreasing $h$ progressively shifts the stain from a standard coffee ring (control) to a more uniform coffee disk ($h = 3 \, {\rm mm}$) to a central deposit with an increasingly smaller inner diameter ($h = 2.5 \, {\rm mm}$ and $h = 2 \, {\rm mm}$). Interestingly, in the latter cases, because of the strengthening of the flows, the deposit also changes from a monolayer ($h = 2.5 \, {\rm mm}$) to a multilayer ($h = 2 \, {\rm mm}$) of particles. For smaller values of $h$, however, even stronger flows produce asymmetric jets that recirculate the suspended colloids away from the center through the top of the droplet (Supporting Movie 4), thus disrupting the symmetry of the final stain. Overall, these patterns are the result of the interplay between the Marangoni flows and the sedimentation velocity $v_{\rm s}$ of the microparticles: for $h \le 2 \, {\rm mm}$, the flows are the main responsible for the particle's deposition dynamics as the sedimentation velocity $v_{\rm s}$ is mostly negligible compared to their intensity; for $h > 2 \, {\rm mm}$, however, as $v_{\rm s}$ and $v_{\rm r}$ become more comparable, sedimentation starts to play a more tangible role, thus contributing to the formation of larger more uniform deposits, such as monolayers and coffee disks. 
 
As shown in Fig.~\ref{fig2}(c), the order of magnitude of these flows as a function of $h$ can be estimated using Eq.~\ref{eq2} as \cite{Nikolov2002}:
\begin{equation}\label{eq3}
v(h) = \frac{h_{\rm D}}{2 \eta R} [ \gamma(h,R) - \gamma(h,0)  ], 
\end{equation}
where $\eta$ is the viscosity of water. The analytical values for $v(h)$ offer an upper bound estimate of the experimental values, as they are in fact more comparable with the average maximum radial velocities recorded during evaporation. In particular, the analytical functional form reproduces reasonably well that of the experimental data for $h \geq 2\, {\rm mm}$. Below this threshold value instead, we can observe a slight deviation between experiments and model. When the vapor point source is too close, the droplet's upper surface is in fact deformed into a doughnut shape due to extra surface tension stress that alters the velocity profile within the droplet (Supporting Movie 2), as also confirmed by the flow velocity maps in Fig. \ref{fig2}a, where a region of near-zero velocity appears at $r = 0$ for $h \leq 2 \, {\rm mm}$. For $h \leq 1 \, {\rm mm}$, this region already appears in the initial phase of the droplet's evaporation and quickly spreads outwards with time. Figure \ref{fig2}(a) also shows how the maximum radial flow velocity is reached after an initial transient caused by ethanol accumulation in the droplet over time as a consequence of its recirculation by the Marangoni flows. The duration of this transient depends on the proximity of the vapor point source to the droplet's surface: the further the needle, the longer this transient.

\begin{figure}
\includegraphics[width=\linewidth]{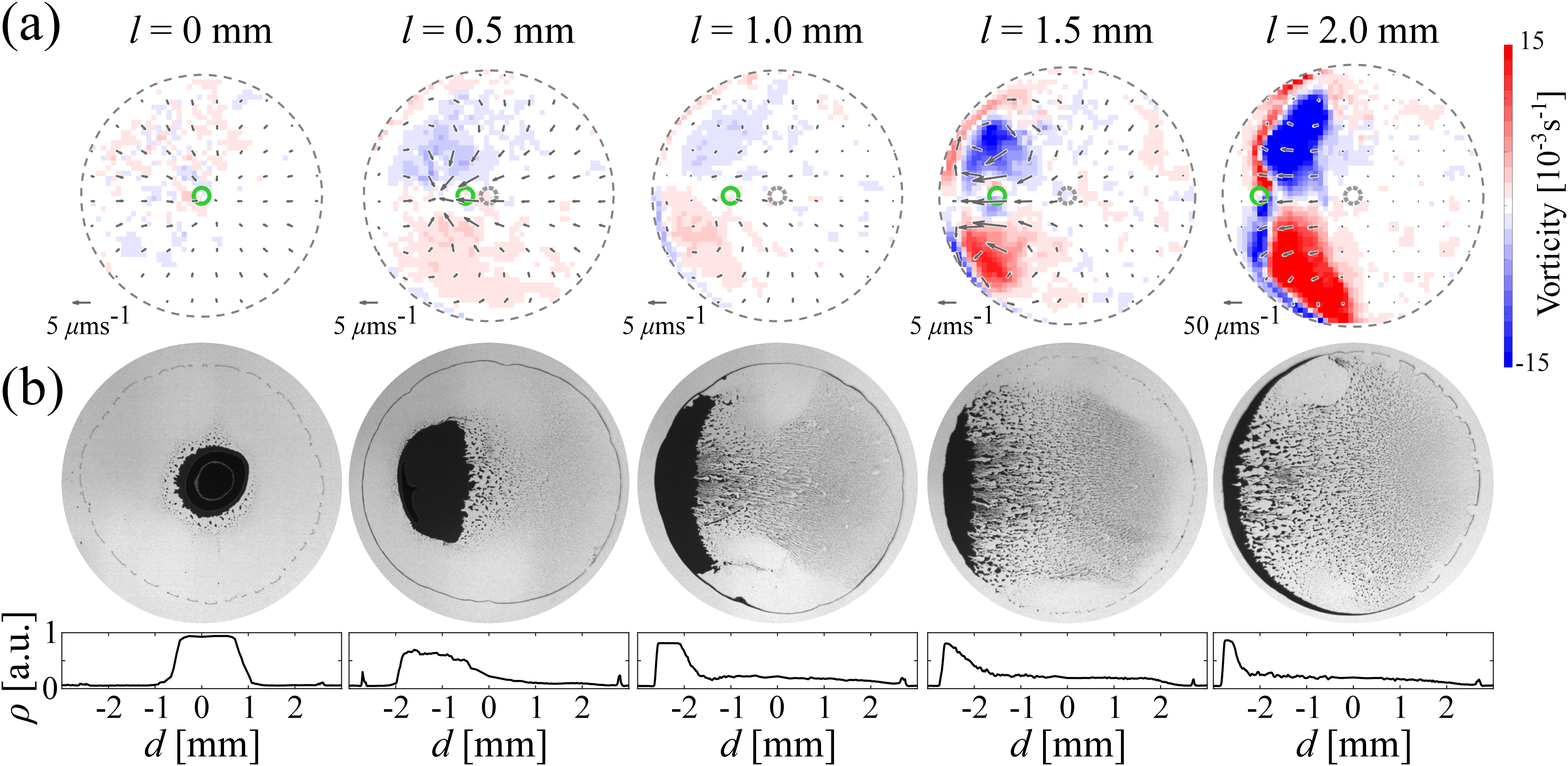}
\caption{\label{fig3} \textbf{Spatial control of the Marangoni flows by the later displacement of the vapor point source.} (a) Flow field maps showing both velocity vectors and vorticity as a function of the lateral displacement $l$ of the vapor point source from the droplet's center for $h = 2 \, {\rm mm}$. The solid circle indicates the position of the needle, the outer dashed circle indicates the edge of the droplet while the inner dashed circle indicates its center. All maps were obtained $50 \, {\rm s}$ into the droplets' lifetime. (b) Final deposits after evaporation for the droplets whose flow velocity maps are shown in (a). The bottom histograms show the corresponding deposit's density profile $\rho$ (averaged along the angular coordinate) along one droplet's diameter (calculated as in Fig.~\ref{fig1}).}
\end{figure}

Beyond the possibility of modulating their strength with the distance of the point source from the droplet, the emergence of the Marangoni flows can also be controlled in space by laterally offsetting the needle (Fig.~\ref{fig3}), thus shifting the position of the minimum in surface tension. Figure \ref{fig3}(a) shows the vectorial velocity maps, and their relative vorticity around the out-of-plane axis \cite{Christy2011}, for different values of the lateral displacement $l$. The distance of the needle from the substrate was fixed at $h = 2 \, {\rm mm}$ to prevent the formation of asymmetric jets, and all maps were obtained $50 \, {\rm s}$ into the droplets' lifetime after the initial transient part of their evaporation due to ethanol accumulation had elapsed. As already noted in Fig.~\ref{fig2}, for no displacement ($l = 0$), the flow is radially symmetric and pointing inwards to the droplet's center, thus virtually presenting zero vorticity. However, when the vapor point source is displaced towards one of the edges, the flows become radially asymmetric, weakening between the source and the edge and strengthening at the opposite side with increasing values of $l$. Because of the displacement of the flow stagnation point towards the edge, the asymmetric compression of the flow lines due to the confined geometry of the droplet induces vorticity below the point source whose strength increases with $l$. Interestingly, the flow stagnation point is not immediately beneath the needle but closer to the droplet's edge as can also be appreciated by the lateral position of the final deposits in Fig.~\ref{fig3}(b). As a consequence of stronger flows and vorticity near the edge, these patterns form closer to the edge and spread more along it when $l$ increases.

\begin{figure}
\includegraphics[width=.5\linewidth]{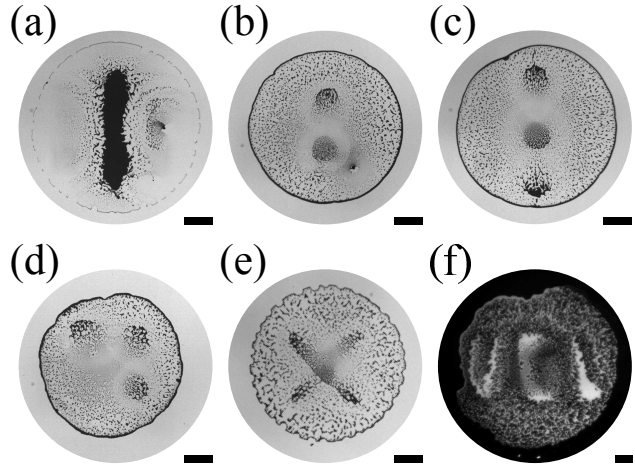}
\caption{\label{fig4} \textbf{Surface patterning based on dynamic spatiotemporal control of the Marangoni flows.} Different surface patterns of deposited colloidal particles after evaporation obtained by spatially shifting the ethanol vapor point source in (a-b) one dimension or (d-f) two dimensions in time: (a) a line, (b) two dots, (c) three dots in line, (d) three dots in a 2D configuration, (e) a cross, and (f) the letters UCL. All patterning was performed using 1-${\rm \mu}$L droplets containing a 1-wt\% water suspension of 2-${\rm \mu}$m monodisperse silica particles, except for (f) where a 3-${\rm \mu}$L droplet of a 1.5-wt\% suspension was used. As vapor point source a needle of 80-${\rm \mu m}$ internal radius at $h \leq 0.75 \, {\rm mm}$ was used, except in (a) where a needle of 210-${\rm \mu m}$ internal radius at $h = 2 \, {\rm mm}$ was used instead (Supporting Table 1). All scale bars correspond to $1\, {\rm mm}$.}
\end{figure}

Finally, after showing control over the strength and the spatial position of the Marangoni flows, we can apply this mechanism to pattern the substrate in a versatile manner by dynamically modulating the emergence of these flows within the droplet in space and time (Fig.~\ref{fig4} and Supporting Table 1). By moving the point source over the droplet's surface during its lifetime, it is possible to dynamically shift the location of the minimum in surface tension, thus allowing for a near-to-real-time reconfiguration of the flows within the droplet. Figure \ref{fig4}(a), for example, shows how a periodic one-dimensional displacement of the point source over the droplet allows to print a line of microparticles (Supporting Table 1). When the needle size is reduced ($r_0 = 80 \, {\rm \mu m}$), it is possible to create more intricate patterns as a result of a higher resolution in the spatial localization of the flows and of the fact that the point source can be brought closer to the droplet's surface without observing the formation of disruptive jets (Supporting Fig. S1): Figures \ref{fig4}(b-c) respectively show the formation of two and three dots in line by sequentially holding the point source at different locations during the evaporation (Supporting Table 1). The addition of a second degree of freedom to the in-plane displacement of the vapor point source allows for patterning the three dots in a 2D configuration (Figs.~\ref{fig4}(d)) and, more broadly, to achieve complex 2D shapes (Supporting Table 1), such as a cross (Fig.~\ref{fig4}(e)) or the letters UCL (Fig.~\ref{fig4}(f)). 

In conclusion, we have proposed a novel experimental mechanism to dynamically control the deposition of particles within an evaporating sessile droplet with an external point source of vapor. In particular, we have demonstrated versatile surface patterning with colloids \cite{Han2012,Larson2013}. Our method relies on the deterministic and dynamic control of the emergence and strength of Marangoni flows within the droplet in space and time. The strength of the flows is controlled by the proximity of the point source to the droplet's surface, while their onset in space and time can be tuned by its lateral offset with respect to the droplet's center. Further control over flow generation could be achieved with different solvents other than ethanol. Differently from other patterning techniques based on the local generation of Marangoni flows \cite{Ta2016,Varanakkottu2016}, our mechanism does not require complex setups, acts without altering the system's temperature and, provided that the solvent'sevaporation point is lower than water, with minimal change in its composition, thus not interfering with the material content of the final deposit. Our results therefore open new avenues in controlling the deposition of patterns and the flow dynamics  within sessile droplets with potential applications in printing, thin-film deposition, self-assembly and the development of diagnostic tools and bioassays.

% \section{Results and discussion}
% \subsection{Outline}

\begin{acknowledgement}

We acknowledge the COST Action MP1305 ``Flowing Matter'' for providing several meeting occasions. Giorgio Volpe acknowledges funding from the HEFCE's Higher Education Innovation Fund (KEI2017-05-07). Giovanni Volpe acknowledges funding from the European Research Council (ERC Starting Grant ComplexSwimmers, grant number 677511). Robert Malinowski and Ivan P. Parkin acknowledge funding from EPSRC (EP/G036675/1). Ivan P. Parkin also acknowledges funding from EPSRC (EP/N510051/1).

\end{acknowledgement}

\begin{suppinfo}

\begin{itemize}

  \item Materials and Methods.
  \item Calculation of the sedimentation velocity $v_{\rm s}$.
  \item Supporting Fig. S1: Effect of the point source's radius on the final deposition pattern. 
  \item Supporting Fig. S2: Spreading of water-ethanol binary mixtures during evaporation on a clean glass slide.
  \item Supporting Movie 1: Standard coffee ring effect. 
  \item Supporting Movie 2: Particle deposition in the presence of an ethanol vapor point-source.
  \item Supporting Movie 3: Evaporation of a water-ethanol binary mixture for ${\rm [EtOH]} = 0.05 \, {\rm v/v}\%$. 
  \item Supporting Movie 4: Radial jets formed in the presence of an ethanol vapor point-source.
  \item Supplementary Table 1: Patterning protocol for Fig.~\ref{fig4}.
  \item Supporting Reference.

\end{itemize}

\end{suppinfo}

\providecommand*\mcitethebibliography{\thebibliography}
\csname @ifundefined\endcsname{endmcitethebibliography}
  {\let\endmcitethebibliography\endthebibliography}{}

\end{document}